# Chapter 17: Mind

**Liane Gabora**
University of British Columbia
liane.gabora@ubc.ca

## Introduction: What Can Archaeology Tell Us about the Mind?

What can relics of the past tell us about the thoughts and beliefs of the people who invented and used them? Recent collaborations at the frontier of archaeology, anthropology, and cognitive science are culminating in speculative but nevertheless increasingly sophisticated efforts to unravel how modern human cognition came about. By considering objects within their archaeological context, we have begun to piece together something of the way of life of people who inhabited particular locales, which in turn reflects their underlying thought processes.

Comparing data between different sites or time periods tells us something about the horizontal (within generation) or vertical (between generations) transmission of material culture. In addition to patterns of transmission of existing kinds of artifacts, we are also interested in novel artifacts that might be indicative of new cognitive abilities, belief structures, or levels of cooperation. An archaeological period marked by the sudden appearance of many kinds of new objects may suggest the onset of enhanced creative abilities. By corroborating archaeological findings with anthropological data (evidence of sudden change in the size or shape of the cranium, for example) with knowledge from cognitive science about how minds function, we can make educated guesses as to what kinds of underlying cognitive changes could be involved, and how the unique abilities of Homo sapiens arose.

In this chapter, we consider three questions about human cognition that can be addressed through archaeological data: (1) How did human culture begin? (2) Where, when, and how did humans acquire the unique cognitive abilities of modern Homo sapiens? and (3) What role do artifacts play in the evolution of these cognitive abilities?

### (1) The Origin of Human Culture

The earliest known artifacts, referred to as Oldowan -- after Olduvai Gorge, Tanzania, where they were first found -- were simple stone tools, pointed at one end (Leakey 1971), and widely associated with Homo habilis, although it is possible that they were also used by late australopithecenes (de Baune 2004). The oldest of these Oldowan tools, from Kada Gona and Kada Hadar, date to approximately 2.5 million years ago (Semaw *et al*. 1997). The earliest use of stone tools was probably to split fruits and nuts; "the moment when a hominin ...produced a cutting tool by using a thrusting percussion ...marks a break between our predecessors and the specifically human" (de Baune 2004:142). With sharp edges, Lower Palaeolithic tools could be used to sharpen wood implements and occasionally butcher small game. Using the sharp flakes associated with these tools, Homo habilis butchered animals for meat, as these tools are found in context with cut-marked bones at the FLK Zinj site (1.75 Ma) in the Olduvai Gorge (Leakey 1971; Bunn and Kroll 1986). Although it has been debated whether they were scavenging, hunting, or "power scavenging" by scaring carnivores away (e.g., Binford 1977; 1983, 1988; Bunn and Kroll 1986; Selvaggio 1998), it is clear that these hominids were capable of acquiring the richer, more meat-rich portions of a kill (Bunn and Kroll 1986). This may have introduced a positive feedback cycle, in that meat eating, by allowing a smaller gut, provided metabolic support for a larger brain (Aiello and Wheeler 1995) which, in turn, would improve cognitive ability for group hunting using mental landscape maps, interpretation of visual clues such as animal tracks, and knowledge of predator behavior.

With the arrival of Homo erectus by 1.8 Ma, we see sophisticated, task-specific stone hand axes, complex stable seasonal habitats, and long-distance hunting strategies involving large game. The size of the Homo erectus brain was approximately 1,000 cc, about 25% larger than that of Homo habilis, and 75% the cranial capacity of modern humans (Aiello 1996; Ruff *et al*. 1997; Lewin, 1999). By 1.6 Ma, Homo erectus had dispersed as far as Southeast Asia (Swisher et al. 1994), indicating the ability to migrate and adapt to vastly different climates (Cachel and Harris 1995; Walker and Leakey 1993; Anton and Swisher 2004). In Africa, West Asia, and Europe, Homo erectus carried the Aschulean handaxe, which was present by 1.4 Ma in Ethiopia (Asfaw et al. 1992). A do-it-all tool that may even have had some function as a social status symbol (Kohn and Mithen 1999) these symmetrical biface tools probably required three stages of production, bifacial knapping, and considerable skill and spatial ability to achieve their final form. The handaxe persisted as a tool of choice for over a million years, spreading by 500 ka into Europe, where was it used by H. heidelbergensis until the Mousterian at about 200 ka. This period also marks the first solid evidence for controlled use of fire, by 800 ka in the Levant (Goren-Inbar et al. 2004). A 400,000 year-old, fire-hardened wooden spear found at Schöningen, Germany (Thieme 1997) shows beyond doubt that Middle Pleistocene Homo were sophisticated big-game hunters (Dennell 1997).

## **Cognitive Theories of How Culture Began**

By a half million years ago the complexity of Homo culture was already beyond that of any other species (Darwin 1871; Plotkin 1988). What got the ball rolling, enabling the process of enculturation to take hold? One suggestion is a theory of mind (Cheney and Seyfarth 1990; Premack and Woodruff 1978), which refers to the capacity to reason about the mental states of others, such as interpreting the intentions of another through his/her actions. However, there is abundant evidence that many species are capable of social attribution and deception, indicating that theory of mind is not unique to humans (Heyes 1998). Moreover, the invention of tools, taming of fire, and conquering of new habitats could not have been achieved solely through enhanced social skills; it would have required enhanced mastery of the physical environment.

Another suggestion is that culture arose due to onset of the capacity for imitation (Dugatkin 2001; Richerson and Boyd 1998). However, imitation is widespread among animals such as dolphins, dogs, primates, and birds (Bonner 1980; Byrne and Russon 1998; Lynch and Baker 1994; Robert 1990; Smith 1977). Moreover there is no evidence, empirical or theoretical, that imitation-like social learning processes (or simple manipulations of them such as a bias to copy high prestige individuals) are sufficient for cultural evolution. To get the kind of adaptive modification and diversification of stable form found in culture (and biology) requires entities that possess an integrated, self-modifying structure wherein changes to one part percolate to other parts with context-sensitive effects, such that different structures can potentially result. Artifacts, habits, and other socially acquired elements of culture do not on their own possess such a structure.

A third proposal is that Homo underwent a transition from an episodic mode of cognitive functioning to a mimetic mode (Donald 1991). The early hominid, functioning in the episodic mode, was sensitive to the significance of episodes, could encode them in memory, and coordinate appropriate responses, but not voluntarily access them independent of environmental cues. Thus awareness was dominated by what was happening in the present moment. Donald maintains that with encephalization (brain enlargement) Homo acquired the capacity for a 'self-triggered recall and rehearsal loop'. which basically amounts to the capacity for a stream of thought in which one idea evokes another which evokes another and so forth recursively. This process, sometimes referred to by psychologists as representational redescription (Karmiloff-Smith, 1992), allowed hominds to direct attention away from the external world toward internal thoughts, planned actions, or episodes encoded in memory, and access them at will. This gave rise to a mimetic mode of cognitive functioning. The term comes from the word 'mime' or 'pantomime', and indeed Donald stresses onset of the ability to act out events that may have occurred in the past or could occur in the future. Thus the mimetic mind is able to not only escape the here and now, but through gesture facilitate a similar temporary escape from the present in other minds. This provided a way of sharing knowledge and experience without language, and may have been our earliest means of cultural transmission. Self-triggered thought also enabled hominids to evaluate and improve skills through repetition or rehearsal.

## **The Underlying Memory Structure: Fine-grained Distributed Representations**

The mimetic mode rests heavily on the capacity to generate previous or imagined episodes or motor acts and adapt them to the present situation. How might it have come about, and why would it be enhanced by encephalization? To answer this we look briefly at the structure of memory. Episodes etched in memory are distributed across a cell assembly (a bundle of brain cells) that contains many locations, and likewise, each location participates in the storage of many items. Thus the same memory locations get used and reused (neural re-entrance). Each memory location is sensitive to a range of subsymbolic microfeatures: primitive attributes of a stimulus or situation. This kind of architecture is said to be content-addressable because similar or related items activate, and get encoded in, overlapping memory regions. Now imagine a hominid observing various features of the landscape. Memory location A may respond preferentially to lines oriented at say 45 degrees from the horizontal, neighboring location B to lines at a slightly different angle, say 46 degrees, and so forth. However, although A responds *maximally* to lines of 45 degrees, it responds to a lesser degree to lines of 46 degrees. This kind of organization is referred to as coarse coding. Thus for example, location A participates in the storage of all those memories involving lines at close to 45 degrees, and a particular episode affects not just A but many other locations. We refer to this constellation of locations as the cognitive receptive field (CRF) for that episode.

As brains size increased, CRFs became larger. This enables more fine-grained encoding of episodes in memory, which makes the memory more prone to self-triggered thought. Consider, for example, an episode that involves getting pricked by a thorny cactus. A small brain might encode this episode in terms of a few features (Figure 1a), such as perhaps the pain involved, and some recognizable feature such as its overall shape or color. This would aid avoidance of contact with this cactus in the future, and perhaps similar cacti. However, that would be the extent of this episode's usefulness. A larger brain, though, might include in the CRF additional features (Figure 1b), such as the sharp point of the thorn. The encoding of an episode constituted by both pointed shape and torn flesh means that there is an association between these properties. Some time later, the need to kill prey may activate the memory locations involved in the encoding of torn flesh, leading to the retrieval of the entire episode, including the pointed shape, and thereby aid in the invention of a spear. Hence the transition from coarse coding, wherein episodes tend to be encoded separately, to fine-grained coding, whereby episodes overlap and can thereby be associated, may have brought the capacity for abstract relationships. Such relationships constitute the stepping stones via which abstract streams of thought occur (through representational redescription), and by which an internal model of the world is acquired and thereafter honed as new information is continually integrated. Abstraction thereby imparts the capacity to reason about anything including, but not limited to, the mental states of others.

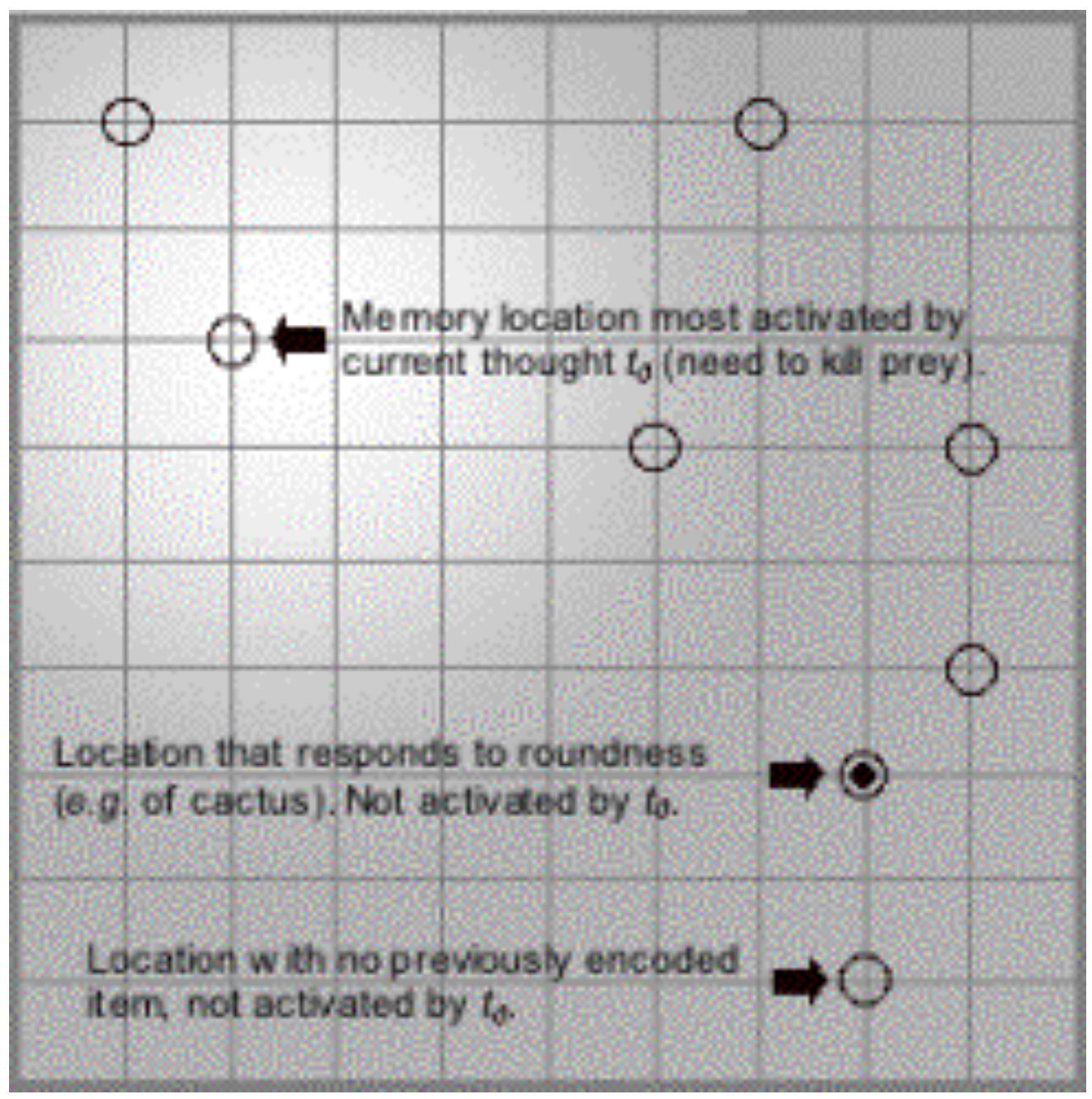

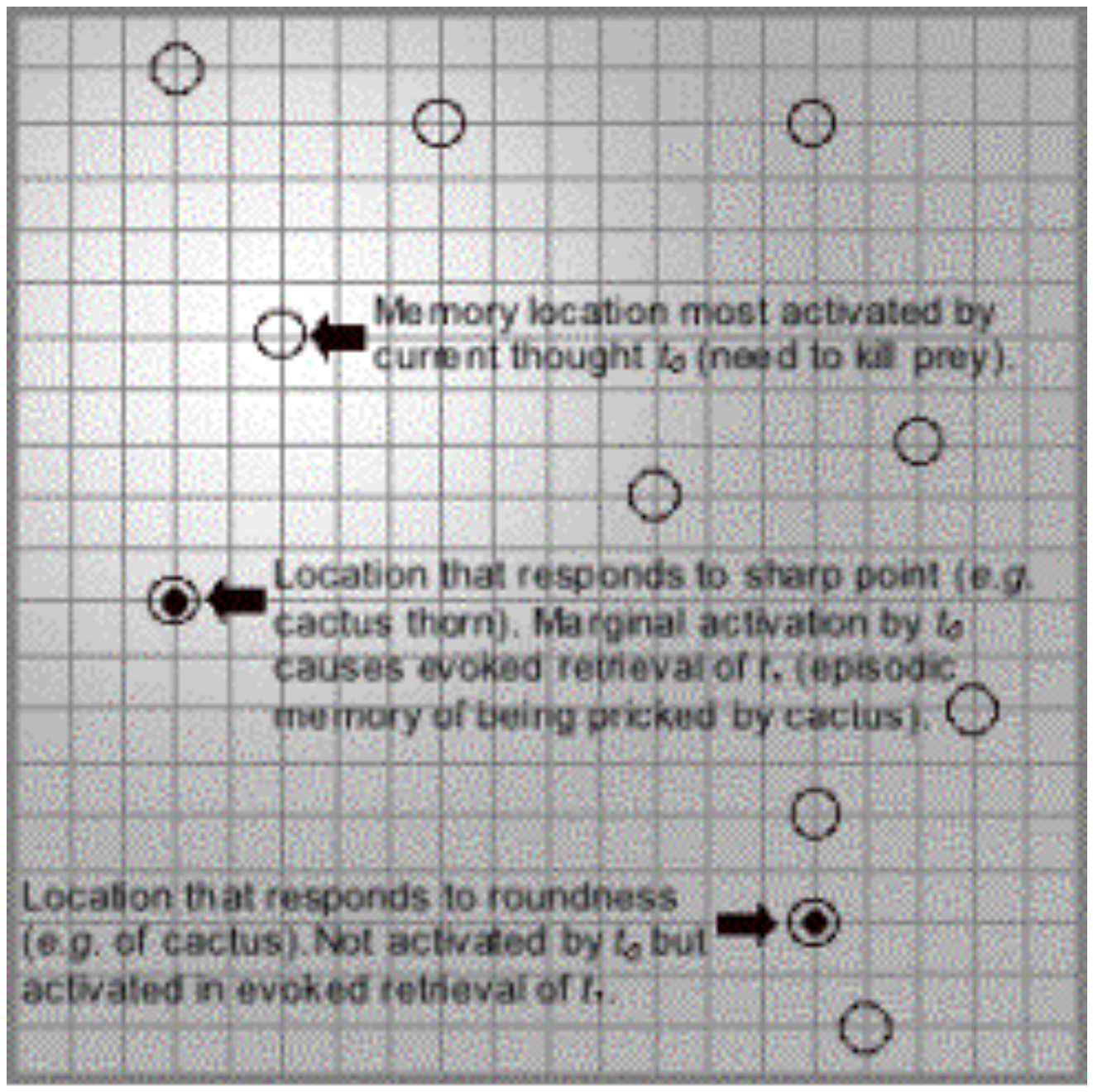

**Figure 1**. (a) Schematic representation of activation of portion of (a) less fine-grained memory, left and (b) more fine-grained memory, right. Each vertex represents a possible memory location. Each black ring represents an actual location. Ring with circle inside represents actual location with previously encoded item. Degree of whiteness indicates degree of activation in current thought, which concerns the need to kill prey for food.

## (2) Cognitive Modernity

The European archaeological record suggests that a profound cultural transition occurred between 60,000 and 30,000 ka at the onset of the Upper Paleolithic (Bar-Yosef 1994; Klein 1989a; Mellars 1973, 1989a, 1989b; Soffer 1994; Stringer and Gamble 1993). The Upper Paleolithic has often been referred to as a "revolution". Considering it "evidence of the modern human mind at work," Richard Leakey (1984:93-94) describes the Upper Palaeolithic as "unlike previous eras, when stasis dominated, ... [with] change being measured in millennia rather than hundreds of millennia." Similarly, Mithen (1996) refers to the Upper Paleaolithic as the 'big bang' of human culture, exhibiting more innovation than in the previous six million years of human evolution. This marks the beginning of a more organized, strategic style of hunting involving specific animals at specific sites, elaborate burial sites indicative of ritual, the colonization of Australia, and replacement of Levallois tool technology by blade cores in the Near East. In Europe, many forms of art appeared, including naturalistic cave paintings of animals, decorated tools and pottery, bone and antler tools with engraved designs, ivory statues of animals and sea shells, and personal decoration such as beads, pendants, and perforated animal teeth, many of which may have indicated social status (White 1989a, 1989b). White (1982:176) also writes of a "total restructuring of social relations." Clearly the Upper Palaeolithic was a period of unprecedented cultural change. Was it strictly a cultural explosion, or a major event in the biological evolution of our species? In either case, how and why did it happen?

## The Upper Palaeolithic as a Cultural Transition

Although the tradition has been to view the Upper Palaeolithic transition as a "revolution," a more appropriate null hypothesis might well be that it was not a "revolution" at all, and that Lower and Middle Palaeolithic symbolism should be reassessed without our modern expectations about iconography (Bednarik 1992). Henshilwood and Marean (2003) argue that many of the diagnostic traits (such as art, burial of the dead, personal ornaments, blade technology, complex hearths, and seasonal hunting in harsh environments) used to identify behavioral modernity are based on the European Palaeolithic record and lack relevance to African Stone Age. In fact, models being proposed for the gradual development of cognitive modernity well before the Upper Palaeolithic transition are substantially supported by archaeological evidence (Bahn 1991; Harrold 1992; Henshilwood and Marean 2003; White 1993; White *et al*. 2003). McBrearty and Brooks (2000), for example, argue that most of the features associated with a rapid transition to behavioral modernity, at 40–50 ka in Europe, are actually found in the African Middle Stone Age tens of thousands of years earlier, including blades and microliths, bone tools, specialized hunting, long distance trade, art and decoration. Some important examples include the Berekhat Ram figurine from Israel (d'Errico and Nowell 2000), and an anthropomorphic figurine of quartzite from the Middle Ascheulian (ca. 400 ka) site of Tan-tan site in Morocco (Bednark 2003).

If modern human behaviors were indeed gradually assembled as early as 250–300 ka, as McBrearty and Brooks (2000) argue, it would push these changes back into alignment with the most recent spurt in human brain enlargement, between 600,000 and 150,000 ka (Aiello 1996; Ruff *et al*. 1997), well before the Upper Paleolithic. Indeed, an earlier cultural transition might be easier to explain if it coincided with an increase in brain size (Bickerton 1990; Mithen 1998).

As Henshilwood and Marean (2003:630) point out, "there is no anatomical evidence (and there may never be any such evidence) for a highly advantageous neurological change after 50,000 years ago." What can be gleaned from skeletal anatomy is that many modern behaviors were at least anotomically possible by earlier dates. For example, the hyoid bone (which connects to the larynx) of a Neanderthal from Kebara Cave (63,000 BP) indicates the anatomically capacity for language (Arensburg et. al. 1989), and the inner-ear structures of Middle Pleistocene humans from the Sierra de Atapuerca (ca. 800 ka) is indicative of the capacity to hear language properly (Martinez et al. 2004). Based on skeletal anatomy, we cannot even rule out the possibility that <u>Homo erectus</u> used some sort of pre-syntactic vocal communication (Wynn 1998).

From DNA evidence it now seems very likely, although not indisputable (Eswaran 2002; Wolpoff et al. 2004), that Neanderthals were a separate lineage and not the ancestors of modern humans (Caramelli et al. 2003; Krings et al. 1997; Ovchinnikov et al. 2000; Tattersall 1998). If Neanderthals possessed cognitive abilities on par with modern humans, it would suggest that their abilities evolved independently during their 500,000 years of separate lineage from humans (Mithen 1996:141). Neanderthals had an average brain size of 1,520 cc (Klein 1989b:272), which is actually slightly larger than the modern human average. In Europe from 130 ka until at least 50 ka but possibly 30 ka (Klein 2003), Neanderthals survived glacial cycles with fluctuations in game populations, climate, and plants. Particularly in their hunting of large mammals (Marean and Kim 1998), their knowledge of cave location, hoof prints, animal behavior, and potential carcass locations was essential to their survival.

The technical intelligence of Neanderthals was considerable; the Levallois technique of tool making requires 5 stages and technical thinking throughout the process, such that a fixed set of rules would be insufficient to do it and even most students today can't learn to make a Levallois point (Mithen 1996). Neanderthal burials indicate that they at least occasionally buried their dead, and that the sick and elderly were cared for (Bahn 1998; Bar-Yosef et al. 1986; Klein 2003; Mithen 1996:135). As for an aesthetic sense, whether Neandethals were capable of art has become a hotly debated issue (e.g., Davidson 1992; D'Errico et al. 2003 and associated comments). Bednarik (1992) compiled a list of published evidence for pre-Upper Palaeolithic symbolic behavior, including the use of ochre or hematite, crystal prisms and fossils, perforated portable objects, engraved or notched bone fragments, and rock art. While some (Davidson 1992; Chase and Dibble 1992) considered this evidence to be spread too sparsely over time and continents to show any convincing pattern, certain individual cases beg explanation, with the higher profile cases including the decorated bone from Bilzingsleben (see Mithen 1996:161), or the musical flute of bone from Divje Babe I cave (D'Errico et al. 1998b).

More recently the debate has focused on whether Neanderthals are responsible for the Chatelperronian industry, which appears to be the early Upper Palaeolithic descendent of the Mousterian industry (Klein 1989b:335). The site of Grotte du Renne, Arcy-sur-Cure, dated at about 34,000 BP, where a Neanderthal temporal bone (Hublin et al. 1996) is found near Chatelperronian artifacts, has become central to the debate over Neanderthal's cognitive capacity for art and symbolism. Based on arguments of the archaeological context within the Grotte du Renne, d'Errico et al. (1998a, 2003) argued that the Chatelperronian ivory beads, pierced bear incisors and wolf canines, and ivory rings were made by Neanderthals, and made distinctively enough so as to not be mere imitations of Aurignacian ornaments made by contemporary modern humans. The Chatelperronian predates the oldest Aurignacian (both stratigraphically and in $^{14}C$) almost everywhere they are found (d'Errico et al. 2003; Klein 2003) and that the Chatelperronian lithics show stylistic continuity from the preceding Neanderthal Mousterian. If their argument is true, and the Neanderthals really did make simple bone tools, decorate themselves with perforated animal teeth, and put red ochre on their living floors, then their considerable cognitive capacities would require a reexamination of the causes of cognitive modernity (d'Errico et al. 2003; Mellars 1998b).

It has been hypothesized that modernity arose at different times and places (Bahn 1998; Mellars 1998b), though this has become exceedingly unlikely on the basis of recent genetic evidence (Prugnolle et al. 2005). Another problem with this scenario is the apparently extraordinary coincidence that these Neanderthal behaviors would have begun just as modern humans were spreading into Europe (Demars 1998; Hublin 1998; Mellars 1998a; 1998b; Toscano 1998). If Neanderthals survived in southern Iberia for 5,000 - 10,000 years after the arrival of anatomically modern humans in northern Spain, it seems likely that they imitated the Aurignacian objects of the newcomers, possibly with some (intentional or unintentional) modification (Klein 2003). However, D'Errico et al. (2003) counter that the Iberian Neanderthals maintained their traditional culture long after modern humans arrived, indicating they were not simply acculturated through contact.

Even if we accept biological ability for "modern" behaviors developed gradually over hundreds of thousands of years, the question remains about why the development in these behaviors accelerated at or after 60,000 BP. What is perhaps most impressive about this period is not the novelty of any particular artifact but that the overall pattern of cultural change is cumulative; more recent artifacts resemble older ones but have modifications that enhance their appearance or functionality. This is referred to as the ratchet effect (Tomasello 1993), and it appears to be uniquely human (Donald 1998). It may have involved an underlying biological change, as discussed in below. Note however that many dramatic cultural revolutions, such as the Holocene transition to agriculture or the modern Industrial Revolution, occurred long after the biological changes that made them cognitively possible. Thus all that can be said for certain is that by somewhere within the vicinity of 50,000 or more years ago, the hominid mind had not only had acquired the potential for cognitive modernity, but its environment (social, cultural, and physical) enabled it to capitalize on or actualize that potential.

## The Upper Paleolithic as an Event in the Biological Evolution of Cognition

The traditional and currently dominant view is that modern behavior appeared in Africa between 50,000 and 40,000 years ago and then spread out of Africa from anatomically modern humans who, through their biologically evolved cognitive advantages, replaced the pre-existing species, including the Neanderthals in Europe (e.g., Ambrose 1998; Gamble 1994; Klein 2003; Stringer and Gamble 1993). If so, what were these cognitive advantages, and how did they come together in the modern mind? Let us now review some explanations.

**Advent of syntactic language in modern humans**

While earlier Homo and even Neanderthals may have been capable of primitive language, the syntactic aspects appear to have emerged at the start of the Upper Palaeolithic (Aiello and Dunbar 1993; Bickerton 1990, 1996; Dunbar 1993, 1996). Syntax enabled language to become general-purpose, put to use in all kinds of situations, whereas previously it had been reserved for social situations. Carstairs-McCarthy (1999) presents a modified version of this proposal, suggesting that although some form of syntax was present in the earliest languages, most of the later elaboration, including recursive embedding of syntactic structure, emerged in the Upper Paleolithic. Another proposal is that at this time humans underwent a transition from a predominantly gestural to vocal form of communication (Corballis 2002). It is indeed possible that the syntactic elements of language emerged at this time, and language can transform cognition (Tomasello 1999). However, we are still left with the question of what made possible the kind of complex thought processes that sophisticated language requires. Furthermore, we may never know exactly when syntactic language began, due the ambiguity of the archaeological evidence. (Bednarik (1992:30), for example, refer to Davidson and Noble's (1989) argument on the cognitive origins of language and iconic graphic art "a house of cards" with no supporting evidence.)

Connection of domain-specific modules: The notion that the mind is modular refers to the view that many cognitive functions are served by innately channeled domain-specific systems, whose operations are largely independent of and inaccessible to the rest of the mind (Fodor, 1983). It has been suggested that what might appear to be unequivocal evidence of modernity might merely reflect the evolution of cognitive competence restricted to particular domains governed by particular modules (e.g. Renfrew and Zubrow, 1994). Another proposal is that modern cognition arose through the connecting of domain-specific brain modules, enabling cross-domain thinking, and particularly analogies and metaphors (Gardner, 1983; Rozin, 1976). Mithen (1996) suggests that in the Upper Paleolithic, modules specialized to cope with domains such as natural history, technology, and social processes became connected, giving rise to conceptual fluidity. One problem with this proposal is that although there is evidence that certain abilities are handled by different modules (e.g. different brain areas specialized for, say, vision and speech), there is no evidence that different modules handle different domains of life e.g. no religion module or tool module. Moreover, the logistics of physically connecting these modules (whose positions in the brain evolved without foreknowledge that they should one day become connected) would be formidable.

Sperber's (1994) solution is that the modules got *connected* not directly, but indirectly, by way of a special module, the 'module of meta-representation' or MMR, which contains 'concepts of concepts'. Sperber's proposal is consistent with there being more general purpose or 'association cortex' by the onset of the Upper Palaeolithic, although the timing of this cultural transition does not coincide with the increase in cranium size.

Symbolic Reasoning: Deacon (1997) has suggested that the Upper Paleaolithic 'revolution' was due to the emergence of an ability to represent complex, internally coherent abstract systems of meaning, consisting of representations and their properties and relationships. Some of these are grounded in concrete experience, but many are instead processes of systematic abstraction or symbolic reasoning. Thus Deacon's account focuses on the ability to systematically reason with abstract symbols and thereby derive causal relationships. It does not deal with the intuitive, analogical, associative processes through which we unearth relationships of correlation.

Cognitive Fluidity: In contrast, Mithen (1998) proposes that the cultural transition of the Middle/Upper Paleolithic reflects enhanced cognitive fluidity, resulting in the ability to map, explore, and transform conceptual spaces. Mithen refers to Boden's (1990) definition of a conceptual space as a 'style of thinking—in music, sculpture, choreography, chemistry, *etc*.' As for why hominids suddenly became good at transforming conceptual spaces, he is rather vague:

> There is unlikely to be one single change in the human mind that enabled conceptual spaces to become explored and transformed. Although creative thinking seems to appear suddenly in human evolution, its cognitive basis had a long evolutionary history during which the three foundations evolved on largely an independent basis: a theory of mind, a capacity for language, and a complex material culture. After 50,000 years ago, these came to form the potent ingredients of a cognitive/social/material mix that did indeed lead to a creative explosion (Mithen 1998:186).

Although the capacity for a theory of mind, language, and complex artifacts could conceivably have arisen at different times, it is hard to imagine how each could have evolved independently. It is possible they have a common cause fact; a single, small change within a complex interconnected system can have enormous consequences (Bak et al. 1988; see also chapter 15 of this volume by Bentley and Maschner). The possibility that the Upper Paleolithic revolution was triggered by a small cognitive change is consistent with scientific understanding of abrupt phase transitions and with the sudden 'breakthroughs' that appear in the archeological record back to the Lower Palaeolithic (de Baune 2004; Tattersall 2003). However, Mithen's basic point—that it is related to the capacity to explore and transform conceptual spaces—holds promise. A similar proposal is that it was due to onset of the capacity to blend concepts, which greatly facilitated the weaving of experiences into stories or parables (Fauconnier and Turner 2002). The question we are left with is: what sort of change in cognitive functioning would enhance the blending of concepts and exploration of conceptual spaces?

Contextual focus: The above proposals for what kind of cognitive change could have led to the Upper Paleolithic transition stress different aspects of cognitive modernity.

Acknowledging a possible seed of truth in each of them we begin to converge toward a common (if more complex) view. Conceptual blending is characteristic of associative thought, which is quite different from the logical, analytic thought stressed by Deacon. The modern human mind excels at both. Hence the Paleolithic transition may reflect a fine-tuning of the genetics underlying the capacity to subconsciously shift between these modes, depending on the situation, by varying the specificity of the activated cognitive receptive field (Gabora 2003). This is referred to as contextual focus*1[2]* because it requires the ability to focus or defocus attention in response to the situation or context. Defocused attention, by diffusely activating a diversity of memory locations, is conducive to associative thought; obscure (but potentially relevant) properties of the situation thus come into play (Figure 2a). Focused attention is conducive to analytic thought because memory activation is constrained enough to zero-in and specifically operate on the most defining properties (Figure 2b). Thus in an analytic mode of thought, the concept 'giant' might only activate the specific concept of bigness, whereas in an associative mode of thought, the giants of various fairytales might come to mind. Once it was possible to shift between these modes of thought, cognitive processes requiring either analytic thought (e.g. mathematical derivation), associative thought (e.g. poetry) or both (e.g., technological invention) could be carried out more effectively.

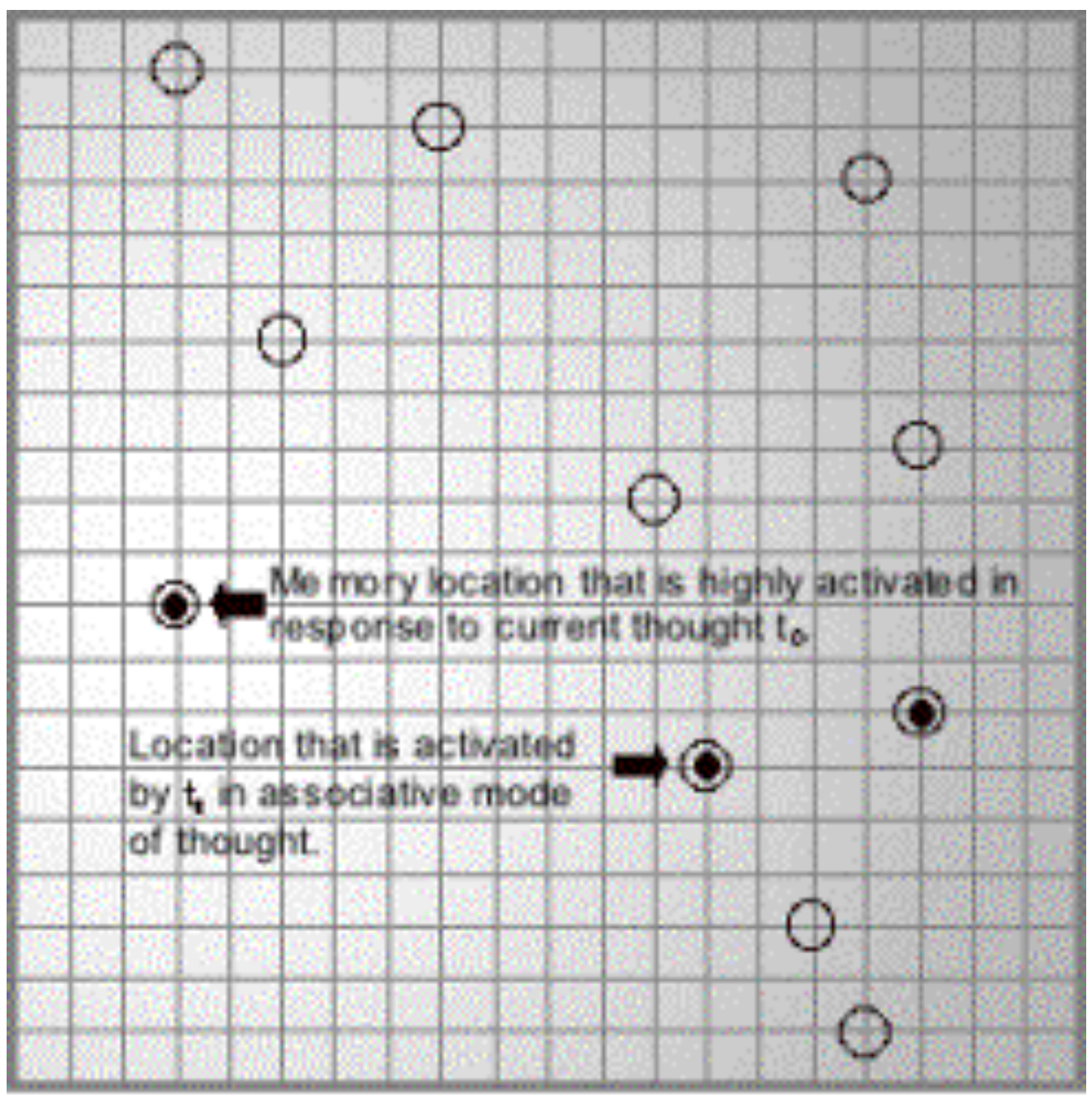

1 In neural net terms, contextual focus amounts to the capacity to spontaneously and subconsciously vary the shape of the activation function, flat for associative thought and spiky for analytical.

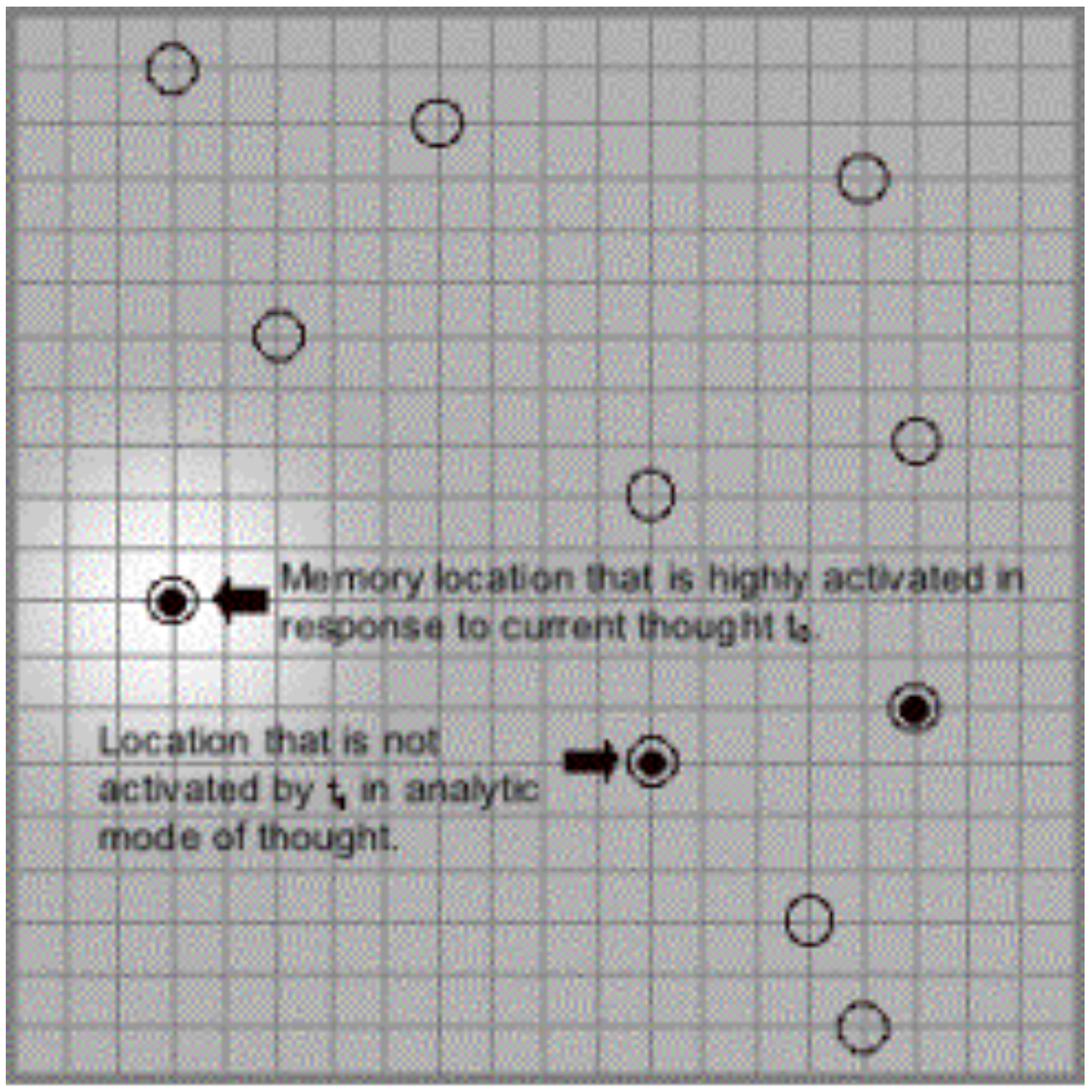

**Figure 2**. (a) Schematic representation of region of memory activated and retrieved from during (a) associative thought, left and (b) analytic thought, right. The activation function is flat in associative thought and spiky in analytic thought. Encoded items in memory locations in whitened region blend to generate next instant of thought.

When we take into account how memory works, the lag between the second rapid increase in brain size approximately 500,000 years ago, and the cognitive transition of the Upper Paleolithic begins to make sense. A larger brain provided more space for episodes to be encoded, but it doesn't follow that this increased space could straightaway be optimally navigated. There is no reason to expect that information from different domains would immediately be compatible enough to coexist in a stream of thought, as in the production of a metaphor. It is reasonable that once the brain achieved sufficient capacity for modern cognition, it took time to fine-tune how its components 'talk' to each other such that different items could be blended together and recursively revised (and recoded) in a coordinated manner. Only then could the full potential of the large brain be realized. Thus the bottleneck may not have been sufficient brain size but sufficient sophistication in the use of the memory already available, which could have come about through onset of contextual focus. With it the modern mind could both (1) intuitively combine concepts, and adapt old concepts to new contexts, and (2) logically analyze these new representations and how they fit together. Together these abilities would lead to the development of more fine-grained internal model of the world, which through language could be expressed and revised. Hence it may not be language per se that made the difference, as Bickerton (1990, 1996) maintained, but rather capacity to shift between analytic and associative processing --- such that the fruits of one mode provide the ingredients for the other --- that not only made language possible, but other aspects of the cultural life of modern humans such as art, science, and religion as well.

## (3) Minds, Artifacts, and Evolution

The cumulative nature of artifact change since the Middle Paleolithic tempts one to refer to culture as an evolution process. Like organisms, artifacts become more complex, more finely honed, and more adapted to the constraints and affordances they impose on one another over time. They also create niches for one another; for example, the invention of cars created niches for seatbelts and gas stations. However, artifacts do not evolve by means of a genetic code, and do not change through a process of random mutation and natural selection. So culture does not evolve in the same sense as biological organisms. In this section we examine in what sense the word evolution applies to artifact change.

## Evolutionary Archaeology

As discussed by Bentley and Maschner in chapter 10 of this volume, "evolutionary archaeology" (EA) is one school among an array of evolutionary approaches that attempt to analyze culture change in terms of concepts borrowed from population genetics such as selection or drift -- changes in the relative frequencies of variants through random sampling from a finite population (O'Brien 1996, 2005; O'Brien and Lyman; for critiques see Boone and Smith, 1998; Gabora, 2006). EA distinguishes itself from other evolutionary approaches by its strong maintenance that artifacts constitute part of an organism's phenotype (defined in biology as the physical manifestation of an organism's genotype) and should be subjected to the same kind of analysis as other phenotypic traits. O'Brien and Lyman (2004: 179) see human artifacts as phenotypic in the biological sense, analogous to "a bird's next, a beaver's dam, or a chimpanzee's twig tools as phenotypic traits." However, their assumption, that heritable behavior is the major determinant in the form of a human-made artifact, is rather easily falsified. Colin Renfrew (1982) succinctly characterized the difference between adaptations that are accumulated genetically and adaptations that are passed on culturally:

> The genetic inheritance, itself the product of environmental selection, may itself be regarded as environmental 'memory'... humans [are not] alone in communicating [their experience of the environment] to others of the species: the dance of the bees, where individuals can indicate the direction and distance of suitable supplies of pollen, fulfils precisely this function. The essence of human culture, however, is that it is an 'acquired characteristic', which, unlike the dance of the bees, is not genetically determined (Renfrew 1982:18).

As Renfrew points out, much of human culture is not genetically determined, meaning that it can evolve in a Lamarckian fashion by inheriting acquired characteristics. Beaver dams and bird's nests, however, cannot evolve this way: if a particular beaver were taught, perhaps by humans, to build a window in its dam, and its offspring still built normal windowless dams, we would see that only the dam, and not the window, belongs in the beaver's biological phenotype, to be passed on via genetic inheritance.

## Do Artifacts Evolve?

Culture has many attributes of an evolutionary process. That is, phenomena such as transmission and drift can operate on cultural variants themselves regardless of the Darwinian (genetic) evolution among humans who carry these variants. In computer simulations (see Costopolous, chapter 16 of this volume), culture evolves through drift or a combination of invention and imitation, without genomes at all (e.g., Gabora 1995; Neiman 1995; Madsen et al. 1999; Bentley et al. 2004). But although culture has attributes of an evolutionary process, it does not operate through the same underlying mechanisms as biology. When it comes to human-made artifacts, inheritance of acquired characteristics is not just the exception but the rule. Once we learned how to build dwellings with windows, dwellings with windows were here to stay. Something is happening through culture that is prohibited in biology.

Some cultural theorists assume that the basic units of this second evolution process are mental representations or ideas, or the tangible products that result from them such as artifacts, language and so forth (Aunger, 2000; Durham, 1991; Lake, 1998). Sometimes these elements of culture are referred to as 'memes', a term that usually implies that they are not only basic elements of culture but also replicators: entities that make copies of themselves (Dawkins 1976). What does that entail? The mathematician and computer scientist John von Neumann (1966) postulated that a replicator, or 'self-replicating automaton,' could be code consisting of (1) a description of itself, or self-description, and (2) a set of self-assembly instructions. After it is passively copied from one replicant, the self-description is actively deciphered via the self-assembly instructions to build the next replicant. In this case, the code functions as interpreted information. Interpreting the code makes a body, while the uninterpreted use of the code makes something that can itself make a body. In biology, the DNA code is copied—without interpretation—to produce new strands of DNA during meiosis. In successful gametes these DNA strands are decoded—interpreted—to synthesize the proteins necessary to construct a body. However, an artifact (or idea) is not a replicator because it does not consist of self-assembly instructions. It may retain structure as it passes from one individual to another. But it does not replicate it. Like a broadcast signal received by a radio, an idea has no self-assembly code and needs external apparatus to be replicated – it cannot self-replicate in the biological sense. Thus the meme perspective is widely believed to be flawed.

Essentially assuming that von Neumann's definition holds for cultural information, Lake (1998) differentiates between its expression versus its symbolically coded representation. Speaking aloud or singing, for example, both express what is represented by a text or written musical score. Singing expresses what can be represented by written music. However neither expression nor representation is equivalent to interpretation or (uninterpreted) copying of a self-assembly code. A musical score does not, on its own, produce little copies of itself. Symbolic coding is not enough to be a replicator; there must be a coded representation of the self. In sum then, artifacts are not replicators and not the basic unit of an evolutionary process.

## Mind as primitive replicator

This does not mean however that nothing in culture is a replicator. Another possibility is that the modern mind itself constitutes a replicator. Not a von Neumann replicator, which replicates with high fidelity using self-assembly instructions as described above, but a primitive replicator (Gabora 2004), which replicates in an autopoietic sense.An autopoietic structure is one in which the parts reconstitute one another and thus the structure as a whole is self-mending or self-replicating. For example, Kauffman (1993) proposed that the earliest life forms were autopoietic; that is, that pre-DNA life initially replicated through a process in which simple molecules each catalyze the reaction that generates some other molecule in the set and thus as a whole they self-replicate. The proposal arose in response to the well-known 'chicken and egg' problem: which came first, the nucleotides that make up a genetic self-assembly code which through transcription and translation leads to proteins, or the proteins that are necessary to many stages of the transcription/translation process? His answer is neither; a sloppy form of self-replication is possible without any code.

Kauffman's proposal draws upon the topological notion of a closure space: a set of points sufficiently connected such that it is possible to get from any one point to another by following the edges (connections) between them. Closure can be visualized as follows. Imagine you spill a jar full of buttons on the floor. You tie two randomly chosen buttons with a thread, and repeat this again and again. Occasionally you lift a button and see how many connected buttons get lifted, and you find that clusters start to emerge. When the ratio of strings to buttons reaches about 0.5, you pass a threshold where, suddenly, a giant cluster of connected buttons forms, containing most of the buttons. The giant cluster is a closure space because any two buttons within it are connected through strings and other buttons. With primitive replicators the chicken-and-egg problem is solved because self-replication occurs through happenstance interactions of molecules more primitive than either DNA or proteins. Since replication proceeds not through following a code but through happenstance interactions, inheritance of acquired characteristics is possible, and replication has a low fidelity because of it.

Similarly, it is proposed that the enculturated modern human mind constitutes a primitive replicator that emerges through a self-organized process of conceptual closure (Gabora, 2000; Gabora and Aerts 2005). Indeed an analogous paradox arises in considerations of the origin of cultural evolution as arose in consideration of the origin of biological evolution. We know that a mind is not just a collection of imitated cultural elements but an integrated model of how different aspects of experience relate to one another, or worldview. But this leads to the following chicken-and-egg situation: Until memories are woven into a worldview, how can they generate the remindings and associations that constitute a stream of thought? Conversely, until a mind can generate streams of thought, how does it integrate memories into a worldview? How could something composed of complex, mutually dependent parts come to be?

Much as in biology, the chicken-and-egg paradox can be resolved by applying the concept of closure. To apply the concept of closure to cognition, episodes in memory are represented as points (buttons), associative paths between them as edges (strings), and concepts as clusters of connected points. Retrieval of an item from memory evokes another retrieval, which, in turn, evokes yet another. This increases the density of associative paths, and the probability of concept formation. Concepts facilitate streams of thought, which forge connections between more distantly related clusters. The ratio of associative paths to concepts increases until it becomes almost inevitable that a giant cluster emerges and the episodes form a connected closure space. Eventually for any one episode or concept there exists a possible associative path to any other, and together they constitute an integrated conceptual web.

The idea is that, like the self-organized autocatalytic sets postulated to be the earliest forms of life, a mind reach a stage in which it is conceptually integrated when it becomes a primitive replicator. Integrated structure enables it to reason about one thing in terms of another, adapt ideas to new circumstances, frame new experiences in terms of previous ones, or combine information from different domains (as what happens with a joke). It should be stressed that it is not the presence of but the capacity for an integrated worldview that the human species came to possess. An infant may be born predisposed toward conceptual integration, but the process must begin anew in each young mind. The manner in which a worldview replicates is not all-at-once but piecemeal, through social exchange, often mediated by artifacts. Worldviews are therefore not so much replaced by as transformed into (literally) more evolved ones. Like cutting a fruit at different angles exposes different parts of its interior, different situations expose different facets of a mind, and the production of artifacts is one way a mind reveals its current evolutionary state. The process is emergent rather than dictated by self-assembly instructions, and therefore characteristics acquired over a lifetime are heritable. One individual modifies the basic idea of a cup by giving it a flat enough bottom to stay put when not in use, and another individual adds a handle, making it easier to grasp. With each instantiation, the basic idea remains the same but the details change to make it more useful or adapted to specific needs.

Thus while brains were evolving through biological evolution, integrated minds began evolving through cultural evolution. The associative or relational structure of the mind manifests as a capacity to define one element in terms of others, predict how a change to one element will affect another, compare the present state of an idea or artifact with its previous state, or desired state, and thereby refine or improve it. Two necessary steps toward conceptual closure would have been (1) that representations were sufficiently distributed to allow concept formation and self-cued retrieval, and (2) that contextual focus enabled the capacity to shift between associative thought (conducive to bridging domains) and analytical thought (conducive to logical operations within a domain). Thus the notion of conceptual closure fits well with the explanations put forth to underlie the two cultural transitions examined earlier, though at present this is speculative.

## Artifacts and the Extended Mind

In contradiction to the meme perspective, neither a painting nor the ideas that went through the artist's mind while painting it constitute a replicator. Instead, a painting reveals some aspect of the artist's mind (which is a primitive replicator) and thereby affects the minds (other replicators) of those who admire it. As minds become increasingly complex, the artifacts they manifest become increasingly complex, which necessitates even more complex cognitive structure, et cetera. This suggests that artifacts are not the basic unit of either biological or cultural evolution, but that they a crucial if indirect role in the evolution of the integrated mind.

As Renfrew (1982) suggests, artifacts function as external memory storage, which affects both the biological and cultural fitness of their makers and users (see also Donald 1991, 2001). Retrieving information from the depiction of bison on a cave wall or calendrical notches in a log is not so different from retrieving knowledge from memory. Note that elements of the natural world functioned as a form of memory long before there were symbolic artifacts; a look of disapproval on a mother's face could remind a child not to eat a poisonous mushroom as readily as retrieval of a memory of doing this and getting sick. The look on the mother's face is not a material artifact, yet it functions as an external memory source, in much the same way as notches in a log. This corresponds to Clark and Chalmers' (1998) extended mind perspective, in which individual experience is woven together through interaction between the internal and external world, and in which the brain and its social cultural environment co-evolve, mutually provoking change in one another (Deacon, 1997; Durham, 1991; Lumsden and Wilson, 1981).

## Conclusions

There are many ways in which human cognition differs from that of other species, such as our ability to generate and understand complex languages and other symbolic structures appearing in art, science, politics, and religion. The appearance of stone tools (2.5 Ma), strategic big game hunting (1.6 Ma), controlled fire use (0.8 Ma), and complex habitats suggests that prior to the onset of complex language and symbolism, hominid cognition already differed profoundly from that of other species. It has been suggested that this reflects onset of the capacity to imitate, or onset of a theory of mind enabling attribution of mental states to others. However, since these abilities are also present in other social species, a more likely suggestion is that it reflects onset of the capacity for self-triggering of thoughts or motor patterns such that one evokes the next. This underlies abilities ranging from refinement of skills for tool production to enactment of events. A plausible explanation for how it came about is that encephalization enabled mental representations of events and skills to become more broadly distributed and thus the resulting memory was more fine-grained, facilitating the retrieval or reminding events that recursively reiterated constitute a self-triggered thought process.

Many, but not all, archaeologists believe that these distinctly human abilities likely arose about fifty thousand years ago, or perhaps somewhat earlier, during the Middle/Upper Paleolithic. Although encephalization has been occurring throughout the last two million years, a second spurt occurred between 600,000 and 150,000 ka, well before the cultural transition of the Middle/Upper Paleolithic. Thus this second cultural transition cannot be attributed to something so simple as the sudden appearance of a 'language module'; a feasible explanation must involve not new brain parts or increased memory but a more sophisticated way of *using* the available memory. The advent of language or symbol use may be part of the answer, but begs the question what sort of cognitive functioning is capable of producing and using language or symbols. A more in-depth examination of spatiotemporal patterns in the archaeological record, however, reveals not just the pronounced appearance symbolic artifacts, but evidence that they build on one another in cumulative fashion; that is, exhibit the ratchet effect. Indeed some believe this to be the most distinctly human characteristic of all. It is indicative of conceptual fluidity, which involves combining ideas in new ways and adapting old ideas to new circumstances, and requires both the complex mental operations characteristic of analytic thought, and the intuitive, analogical processes characteristic of associative thought. Thus it has been proposed that culture was brought about by onset of the capacity to adapt others' ideas to ones' own circumstances, and this requires (1) an integrated internal model of the world, and (2) a sophisticated way of navigating that model of the world, such that relationships can be retrieved and creatively elaborated. This provides a possible cognitive explanation for the explosion of task-specific decorated tools, beads, pottery, and so forth during the Middle/Upper Paleolithic: they reflect onset of the capacity to spontaneously shift between these two forms of thought depending on the situation. Together, analytical thought allowing calculation within a domain, and associative thought forging connections amongst seemingly disparate domains. This paved the way for episodes encoded in memory to become integrated through the formation of a dynamical network of concepts to yield a self-modifying internal model of the world.

The last section of this chapter looked at the relationship between minds, artifacts, and evolution. Evolutionary archaeologists view artifacts as part of the human phenotype, which increase biological fitness through their functionality. However we saw that although there may be a genetic component to the capacity to produce a well-crafted artifact it is not the only or even dominant influence; culture appears to play a formidable role. We also saw that artifacts themselves do not constitute replicators, and are thus not the basic unit of a cultural evolution process. However it is possible that integrated worldviews evolve in the same primitive sense as the first living organisms, through an emergent, self-organized process, resulting in a structure that can be mathematically described as a closure structure. Because no self-assembly code is involved, the evolution of such emergent structures is Lamarckian; acquired characteristics are inherited.

In this brief chapter it has been possible only to sketch out some ideas about the evolution of human cognitive abilities. We have a long way to go before we arrive at a complete picture, but it is an exciting journey ahead.

**Acknowledgements**
I would like to acknowledgement the support of Foundation for the Future. I would also like to thank Alex Bentley for many helpful comments and additions, and Merlin Donald for comments on portions of this chapter.